\documentclass[journal=jacsat,manuscript=article]{achemso}

\usepackage[version=3]{mhchem} 
\usepackage{tikz}
\usetikzlibrary{intersections}

\usepackage{helvet}

\usetikzlibrary{positioning}
\usetikzlibrary{plotmarks}

\author{Fr\'ed\'eric Peyskens}
\email{fpeysken@intec.ugent.be}
\affiliation{Photonics Research Group, INTEC-department, Ghent University-imec; Center for Nano-and BioPhotonics, Ghent University, Sint-Pietersnieuwstraat 41, 9000 Ghent, Belgium}
\author{Ashim Dhakal}
\affiliation{Photonics Research Group, INTEC-department, Ghent University-imec; Center for Nano-and BioPhotonics, Ghent University, Sint-Pietersnieuwstraat 41, 9000 Ghent, Belgium}
\author{Pol Van Dorpe}
\affiliation{imec, Kapeldreef 75, 3001 Heverlee, Belgium; Department of Physics, KULeuven, Celestijnenlaan 200D, 3001 Leuven, Belgium}
\author{Nicolas Le Thomas}
\affiliation{Photonics Research Group, INTEC-department, Ghent University-imec; Center for Nano-and BioPhotonics, Ghent University, Sint-Pietersnieuwstraat 41, 9000 Ghent, Belgium}
\author{Roel Baets}
\affiliation{Photonics Research Group, INTEC-department, Ghent University-imec; Center for Nano-and BioPhotonics, Ghent University, Sint-Pietersnieuwstraat 41, 9000 Ghent, Belgium}

\title[An \textsf{achemso} demo]
  {Surface enhanced Raman spectroscopy using a single mode nanophotonic-plasmonic platform}
  
\begin{document}


\newpage
\textbf{Surface Enhanced Raman Spectroscopy (SERS) is a well-established technique for enhancing Raman signals. \cite{ref1,ref2,ref3,ref8,ref17,ref18,ref19,ref20,ref21,ref22,ref23,ref24} Recently photonic integrated circuits have been used, as an alternative to microscopy based excitation and collection, to probe SERS signals from external metallic nanoparticles. \cite{ref32,ref33,ref44} However, in order to develop quantitative on-chip SERS sensors, integration of dedicated nanoplasmonic antennas and waveguides\cite{ref25,ref26,ref27,ref28,ref29,ref30,ref31} is desirable. Here we bridge this gap by demonstrating for the first time the generation of SERS signals from integrated bowtie nanoantennas, excited and collected by a single mode waveguide, and rigorously quantify the enhancement process. The guided Raman power generated by a 4-Nitrothiophenol coated bowtie antenna shows an $8\times 10^{6}$ enhancement compared to the free-space Raman scattering. An excellent correspondence is obtained between the theoretically predicted and observed absolute Raman power. This work paves the way towards fully integrated lab-on-a-chip systems where the single mode SERS-probe can be combined with other photonic, fluidic or biological functionalities.}
\par
\begin{figure}[htbp]
  \centering
 	\includegraphics[trim= 1cm 5cm 1cm 5cm, clip=true,width=1\textwidth]{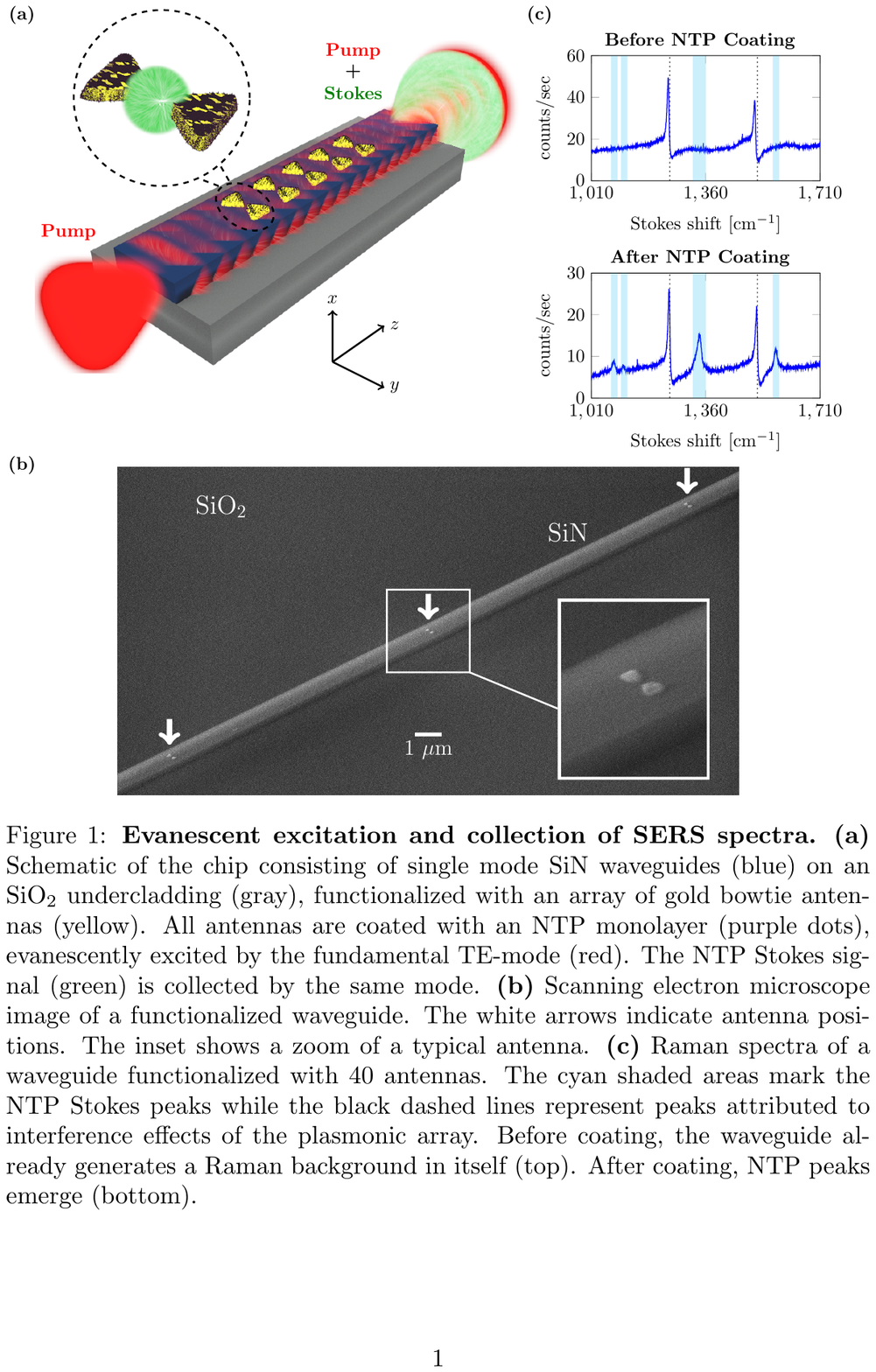} \caption{\label{fig1}}
\end{figure}

A schematic of the device under study is shown in Figure \ref{fig1}(a). The fundamental TE-mode of a silicon nitride (SiN) rib waveguide excites a periodic array of gold bowtie antennas coated with a 4-Nitrothiophenol (NTP) monolayer. The pump wavelength for all experiments is set to $\lambda_{P}=785$ nm and NTP Stokes light (at $\lambda_{S}$) is subsequently collected back into the same waveguide mode. Fabrication details can be found in the Methods section and a description of the measurement setup is outlined in the Supplementary Information S1. A scanning electron microscope image of the functionalized waveguide, with cross-sectional area of $220\times 700$ nm$^{2}$, is depicted in Figure \ref{fig1}(b). Raman spectra of an uncoated and coated waveguide functionalized with 40 antennas are shown in Figure \ref{fig1}(c). The spectral regions where an NTP Stokes peak is expected (1,080, 1,110, 1,340 and 1,575 cm$^{-1}$)\cite{ref41} are highlighted by the cyan shaded areas. Before coating no NTP peaks can be distinguished from the inherent SiN background. The peaks at 1,250 and 1,518 cm$^{-1}$ (marked by the black dashed lines) are attributed to interference effects of the Au array which act on the scattered background light, and they are also observed on the extinction curves of the functionalized waveguides (see Supporting Information S2). Hence they do not represent specific Raman lines. After coating, four additional peaks appear and coincide with the expected NTP Stokes peaks. This demonstrates that SERS signals from single monolayer coated antennas can be efficiently excited and collected by the same fundamental waveguide mode.
\par
\begin{figure}[htbp]
  \centering
 	\includegraphics[trim= 1cm 12cm 1cm 4cm, clip=true,width=1\textwidth]{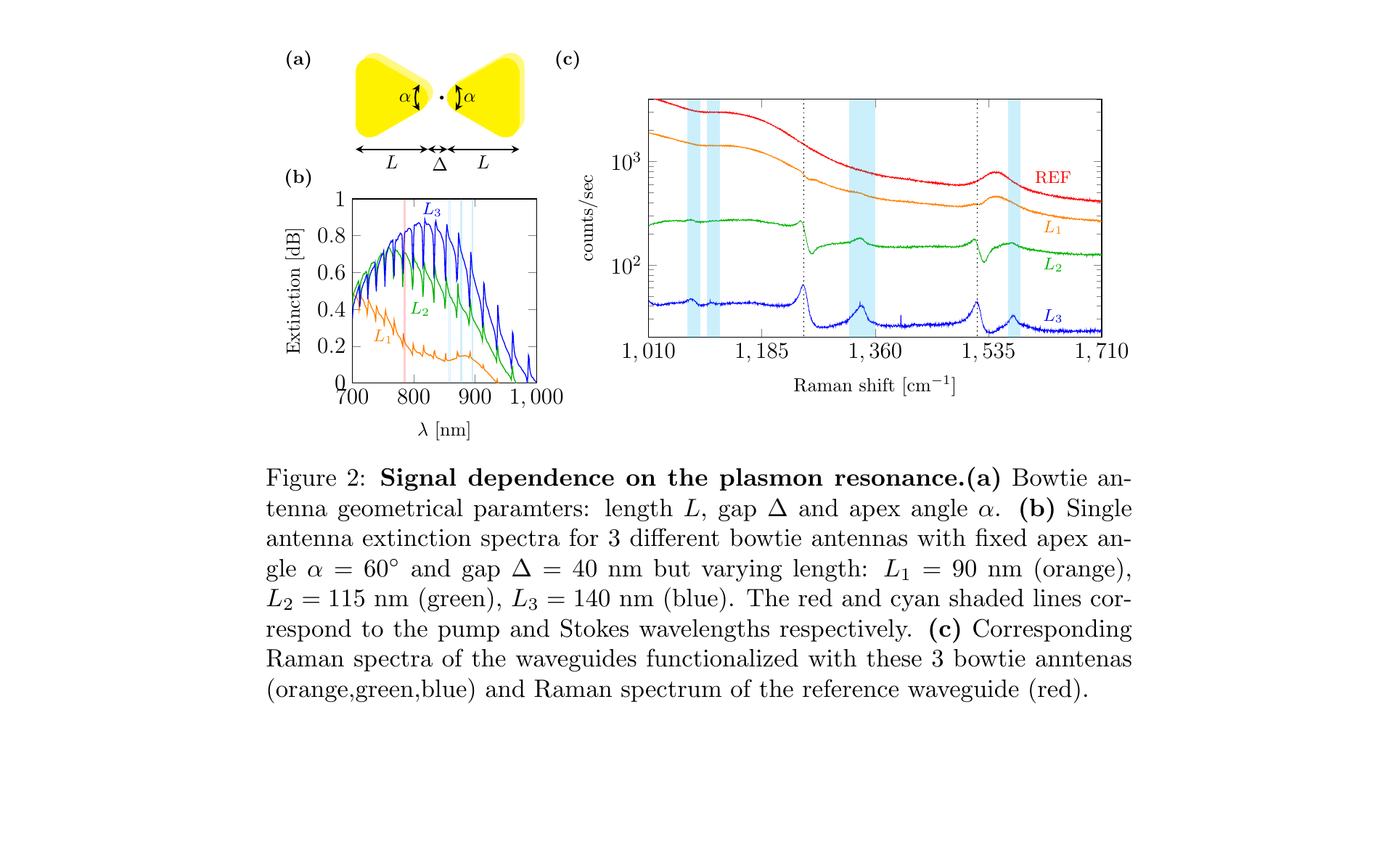} \caption{\label{FigureER}}
\end{figure}

Subsequently the dependence of the SERS signal on the position of the plasmon resonance was investigated to verify that it can be attributed to a resonance effect and not to coincidental surface roughness. To this end, waveguides functionalized with a fixed number of antennas but varying bowtie geometries were considered. The relevant bowtie parameters are its length $L$, gap $\Delta$ and apex angle $\alpha$ (Figure \ref{FigureER}(a)). By changing the length, the antenna resonance can be tuned ($L_{1}= 90$ nm, $L_{2}= 115$ nm and $L_{3}= 140$ nm for fixed $\alpha= 60^{\circ}$ and $\Delta= 40$ nm). Extinction spectra are plotted in Figure \ref{FigureER}(b) while the corresponding Raman spectra are depicted in Figure \ref{FigureER}(c). The Raman spectrum of a reference waveguide without any Au functionalization is also shown. Even after coating the reference waveguide does not generate NTP peaks, so any Raman signal indeed originates from the antenna region and does not contain contributions from spontaneous Raman scattering along the waveguide. \cite{ref36} The $L_{1}$ resonance is detuned from the relevant pump and Stokes region, resulting in a poor Raman spectrum. By increasing the length ($L_{2}$ and $L_{3}$ bowties) the resonance redshifts and lines up with the pump and Stokes wavelengths. For these bowties the NTP spectrum starts to emerge. The reported SERS spectra can hence be attributed to a plasmon resonance effect such that a stable and reproducible enhancement factor can be associated with them, in contrast to SERS events originating from random surface defects. The increased overlap with the plasmon resonance, and hence extinction, also results in a decreased background.
\par
\begin{figure}[htbp]
  \centering
 	\includegraphics[trim= 1cm 15cm 1cm 4cm, clip=true,width=1\textwidth]{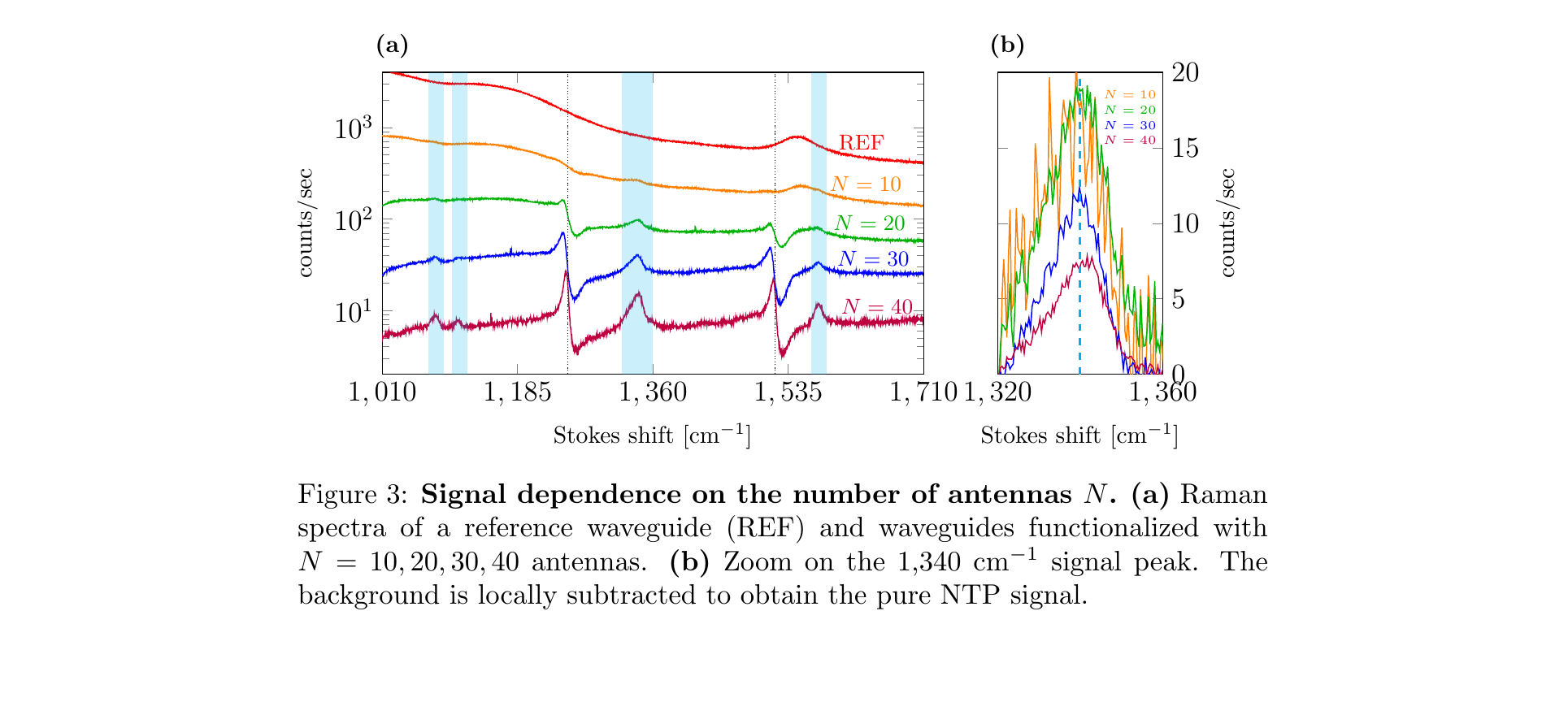} \caption{\label{fig4}}
\end{figure}

Due to the metal induced loss, there will exist an optimum number $N_{opt}$ of patterned antennas such that the SERS signal is maximized. Such an optimum is investigated in Figure \ref{fig4} for a fixed bowtie geometry ($\alpha=60^{\circ}$, $L=100$ nm and $\Delta=40$ nm) but varying $N$: $N=10,20,30,40,70$ and $N=0$ which is a reference waveguide. Each waveguide is measured 10 times and the averaged Raman spectra are reported in Figure \ref{fig4}(a). The $N=70$ signal is not shown since it could not be distinguished from the inherent offset signal of the detector. For a given fixed input power, corresponding to roughly 5 mW guided power, the reference waveguide generates a considerable background signal in the 1,340 cm$^{-1}$ region, where the strongest NTP peak is expected. Functionalizing the waveguide with increasing $N$ reduces this unwanted background due to the attenuation caused by the nanoantennas. In addition the 1,340 cm$^{-1}$ peak starts to emerge when $N$ increases. The smaller peaks at 1,080, 1,110 and 1,575 cm$^{-1}$ only appear when the background is sufficiently low. A zoom on the dominant 1,340 cm$^{-1}$ peak (cyan dashed line) is shown in Figure \ref{fig4}(b). For clarity the background is locally subtracted. As expected, the signal reaches a maximum value for $10\leq N \leq20$ and then decays again with increasing $N$. Apart from signal optimization it is however equally important to reduce the SiN background in order to resolve the smallest spectral features. Therefore an analytical model is developed to outline the interplay between signal enhancement and background reduction, and to derive the relevant figure of merit for on-chip SERS. 
\par
\begin{figure}[htbp]
  \centering
 	\includegraphics[trim= 1cm 10cm 1cm 4cm, clip=true,width=1\textwidth]{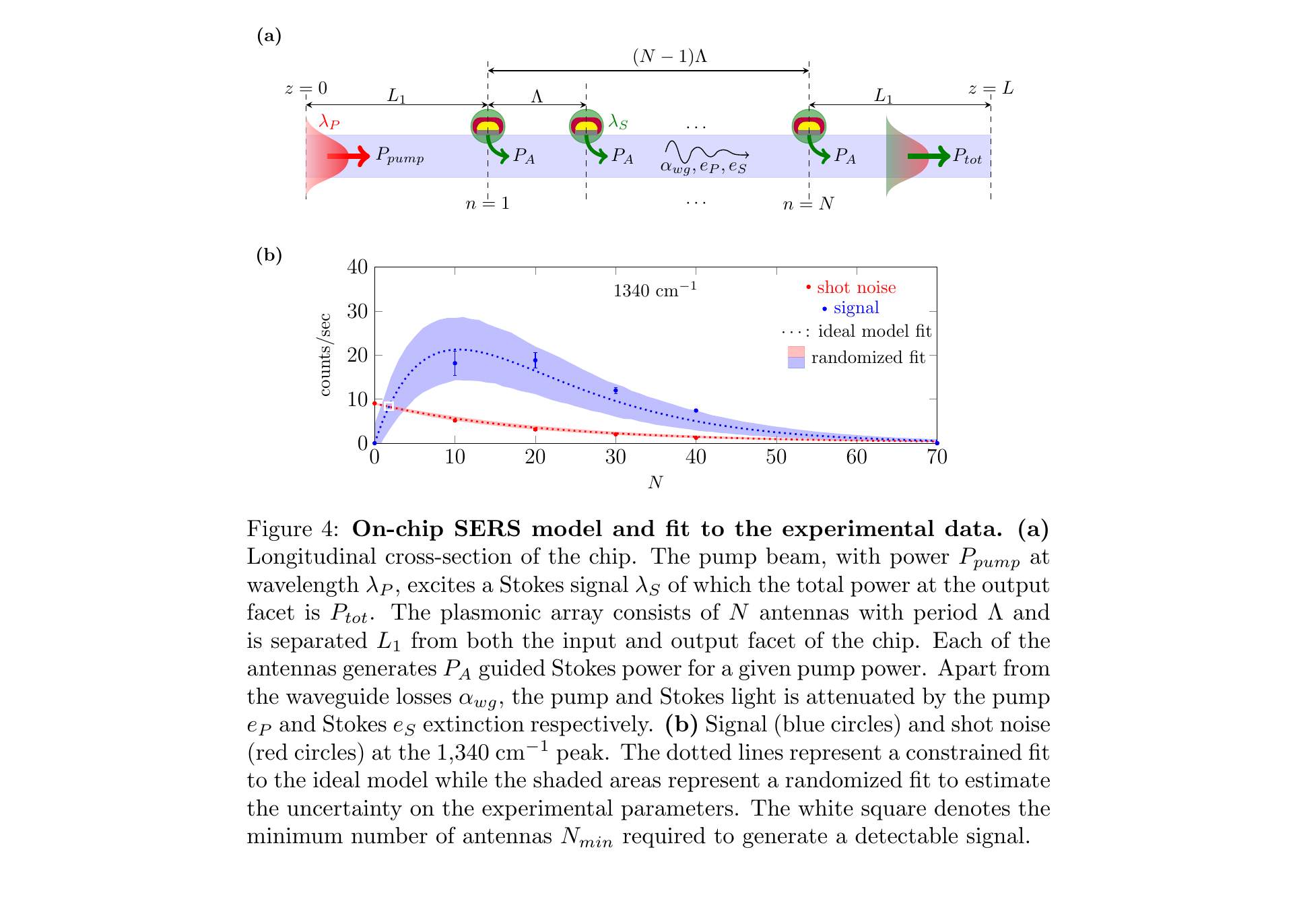} \caption{\label{FigureAAC}}
\end{figure}

Figure \ref{FigureAAC}(a) shows a schematic longitudinal cross-section of the chip consisting of $N$ antennas spaced with period $\Lambda=10\ \mu$m. Each array is centered on the waveguide with a distance $L_{1}\approx 0.5$ cm to the front and back facet of the chip. The NTP monolayer on each antenna will generate a forward propagating Stokes power $P_{A}(\lambda_{P},\lambda_{S})$ for a given pump power $P_{pump}$. This single antenna conversion efficiency $P_{A}(\lambda_{P},\lambda_{S})$ is an antenna dependent factor incorporating the integrated field enhancement profile near the metal surface and the molecular density and Raman cross-section. The total transmission loss induced by one antenna at wavelength $\lambda$ is given by $1-e_{\lambda}^{-1}$, whereby $e_{\lambda}$ is the linear antenna extinction. Apart from the intrinsic waveguide losses $\alpha_{wg}$, the pump and Stokes light will hence also be attenuated by $e_{P}$ and $e_{S}$ respectively. In the Supplementary Information S3 it is then shown that the total Stokes power $P_{tot}$ generated by an array of $N$ coated antennas is approximately given by
\begin{equation}
	\frac{P_{tot}}{P_{pump}}\approx P_{A}(\lambda_{P},\lambda_{S})\text{e}^{-2\alpha_{wg}L_{1}}e_{S}^{\left(1-N\right)}\left(\frac{1-\left(\frac{e_{S}}{e_{P}}\right)^{N}}{1-\left(\frac{e_{S}}{e_{P}}\right)}\right)=FOM(N,\lambda_{P},\lambda_{S})\text{e}^{-2\alpha_{wg}L_{1}}. \nonumber
\end{equation}
The quantity $FOM(N,\lambda_{P},\lambda_{S})$ contains all necessary parameters to assess the SERS signal strength for a given waveguide geometry and is hence considered to be the relevant figure of merit ($FOM$) in comparing the performance of integrated antenna arrays. The optimum antenna number $N_{opt}=\log\left\{\log\left(e_{S}\right)/\log\left(e_{P}\right)\right\}/\log\left(e_{S}/e_{P}\right)$. For the particular bowtie antenna studied in Figure \ref{fig4}, the extinction spectrum $e(\lambda)$ and single antenna conversion efficiency $P_{A}(\lambda_{P},\lambda_{S})$ are numerically evaluated using Lumerical FDTD Solutions (see Methods section). The predicted $N_{opt}$ for the 1,340 cm$^{-1}$ peak is 10 antennas, using the simulated extinctions $e_{P}=1.14$ ($E_{P}=0.58$ dB) and $e_{S}=1.08$ ($E_{S}=0.34$ dB), while $P_{A}\approx 2.35\times 10^{-15}$. For each 1W of pump power the antenna is therefore expected to generate 2.35 fW of guided Stokes power. The Raman enhancement factor is calculated through $EF_{R}=\beta(\lambda_{P})^{2}\beta(\lambda_{S})^{2}$ in which $\beta(\lambda)$ is the local field enhancement (see Methods section). In the center of the gap at 5 nm from the tip of the antenna (marked by the black dot in Figure \ref{FigureER}(a)) an $EF_{R}\approx 1.42\times 10^{4}$ is expected . Apart from the relevant NTP signal, the SiN itself generates a considerable background while the pump beam is propagating along the waveguide. This background signal $P_{bg}$ can be approximated by
\begin{equation}
\frac{P_{bg}}{P_{pump}}\approx P_{B}(\lambda_{P},\lambda_{S})\text{e}^{-2\alpha_{wg}L_{1}}\left(e_{P}^{-N}+e_{S}^{-N}\right) \nonumber
\end{equation}
in which $P_{B}(\lambda_{P},\lambda_{S})$ is a waveguide dependent factor incorporating the specific modal field profile and the SiN molecular density and cross-section (see Supplementary Information S4).
\par
Our analytical model and the associated numerical calculations will now be benchmarked against the spectra from Figure \ref{fig4} to verify whether the theoretically estimated power values correspond to the experimentally obtained absolute Raman powers. To this end, the NTP signal strength at 1,340 cm$^{-1}$, obtained from Figure \ref{fig4}(b), is analyzed as a function of the antenna number $N$ and compared with the theoretical estimations. Furthermore, the background associated shot noise is also calculated. The results are depicted in Figure \ref{FigureAAC}(b). While the ideal model assumes $N$ identical antennas, the fabricated antennas will show differences among each other resulting in changes of $e_{P}$, $e_{S}$ and $P_{A}$ from one antenna to the other. A generalized model incorporating potential differences in $e_{P}$, $e_{S}$ and $P_{A}$ is described in the Supplementary Information S5. In order to estimate the uncertainty on these experimental parameters a randomized fit to the generalized model has been applied. A set of normally distributed numbers is generated for each of the three parameters and then plugged into the generalized model to calculate the distribution of signal and shot noise counts, defining an area within which the probable signal (blue area) and noise (red area) counts are situated. The mean values of these distributions are extracted from an initial constrained fit to the ideal model (dotted lines), while its standard deviations are chosen such that the corresponding signal and shot noise distributions cover all experimental datapoints (red and blue dots). Based on the randomized fit it is possible to estimate the spread on the experimental parameters: $E_{P}\approx 0.49\pm 0.11$ dB, $E_{S}\approx 0.35\pm 0.11$ dB and $P_{A}=(2.60 \pm 0.77) \times 10^{-15}$ (theoretically $E_{P}=0.58$ dB, $E_{S}=0.34$ dB and $P_{A}=2.35\times 10^{-15}$ were predicted). The theoretically predicted parameters are all within the error bars of the experimentally fitted data, which clearly establishes the validity of our model and its ability to provide quantitative predictions of the absolute Raman power coupled into a single mode waveguide. Given this excellent correspondence, we expect the fabricated structures to have a Raman enhancement factor $EF_{R}$ on the order of $1.42\times 10^{4}$ near the two antenna gap tips. Decreasing the gap size should boost $EF_{R}$ and $P_{A}$ by another two or three orders of magnitude. From the fitting values the optimum antenna number is estimated to be $11\pm 3$ (compared to 10 theoretically). The single antenna conversion efficiency $P_{A}$ shows that the fabricated antennas produce $(2.60 \pm 0.77)$ fW of guided Stokes power for each 1W of guided pump power. Compared to the free space Raman scattering $P_{0}$ of a single NTP molecule in a bulk air environment $P_{A}\approx 3.94\times 10^{6}P_{0}$. This includes the excitation and emission enhancement of all molecules in the monolayer as well as the coupling efficiency to the guided mode. Since only half of the Stokes light is carried by the forward propagating mode, the total power coupled into the fundamental TE-mode is therefore $\approx 8\times 10^{6}P_{0}$. 
\par
Our observations also reveal that a minimum number of antennas $N_{min}$ is required to generate a detectable signal (marked by the white square in Figure \ref{FigureAAC}(b)). If $N<N_{min}$ then the shot noise still dominates on the signal. It has to be noted however that the relevant signal is generated in a very small region $(N-1)\Lambda$ compared to the overall length $2L_{1}+(N-1)\Lambda\approx 2L_{1}$, while the shot noise is mainly attributed to this non-useful length $2L_{1}$. Chip designs which allow a separation of the signal from the background are expected to have $N_{min}=1$ such that signals originating from one single antenna can still be detected. As a result it would become possible to simultaneously probe large areas of analytes ($>\lambda^{2}$) and detecting all SERS events, originating from different locations, by monitoring just a single waveguide output in contrast to microscopy based systems where one has to serially scan all hotspot locations.
\par
The work presented here paves the way towards the efficient design of evanescently coupled nanoantennas for on-chip excitation and emission enhancement in the 700-1000 nm region. Due to the low fluorescence, negligible water absorption and the availability of high quality and low-cost sources and detectors this region is of particular interest for Raman sensing.\cite{ref46} In combination with other on-chip spectral functionalities, such as arrayed waveguide gratings\cite{ref46}, the presented platform is forecasted to allow multiplexed detection of extremely weak Raman signals on a highly dense integrated platform. We also envisage that integrated nanoantennas, similar to the ones reported here, could be used as transducer between quantum dot emitters and the fundamental waveguide mode, potentially enabling applications in on-chip quantum communication and quantum computation. \cite{ref43,ref45}

\section{Methods}
\subsection{Fabrication details}
The fabrication consists of a 2-step e-beam lithography process. In the first step the nanoplasmonic antennas are patterned on top of a slab Si/SiO$_{2}$/SiN wafer using a positive PMMA e-beam resist. After PMMA exposure, the samples are developed in a 1:1 MIBK:IPA solution after which a 2 nm Ti adhesion layer and 30 nm Au layer are deposited in a commercial Pfeiffer Spider sputter system. The samples are then immersed in acetone for lift-off. In the second step the waveguides are defined. After metal lift-off a negative ma-N 2403 resist is spun, exposed and developed in ma-D 525. An e-spacer is also spun on top of the ma-N 2403 to avoid charging effects. The developed samples are then etched with an ICP plasma (C$_{4}$F$_{8}$/SF$_{6}$ mixture) in a commercial Oxford Plasmalab system. After resist strip and cleaning, the samples are immersed overnight in a 1 mM NTP:EtOH solution and subsequently rinsed with pure ethanol to remove the residual NTP. A self-assembled monolayer of NTP is assumed to form on the Au surface through a Au-S bond. \cite{ref41}

\subsection{Numerical Simulations}
Numerical simulations were performed with Lumerical FDTD Solutions. We used a refractive index of $n_{rib}=1.9$ for the SiN rib (with width $w_{rib}=700$ nm and height $h_{rib}=220$ nm), $n_{uclad}=1.45$ for the SiO$_{2}$ undercladding and $n_{tclad}=1$ for the top cladding (air). The Si substrate was not taken into account since the real oxide cladding is thick enough to avoid substantial power leakage to the Si such that the numerical results faithfully represent the actual experimental conditions. A thin native oxide layer ($t_{nox}=2$ nm) between the SiN and the Ti has also been incorporated. \cite{ref28} The metal stack thicknesses are fixed to $t_{Ti}=2$ nm and $t_{Au}=30$ nm and a built-in refractive index model for Au (Johnson and Christy \cite{ref37}) and Ti (CRC \cite{ref38}) is used. An additional surface layer with thickness $t_{NTP}=1$ nm and index $n_{NTP}=3$ is used to model the NTP monolayer. The antenna region (including the Ti adhesion layer and the NTP monolayer) is meshed with a uniform mesh of 0.5 nm in the plane of the antenna ($yz$-plane) and 2 nm in the $x$-direction. A mesh refinement to 1 nm is applied in regions where the thicknesses in the $x$-direction are $\leq 2$ nm. The estimated surface area of an NTP molecule is 0.18 nm$^{2}$, so the surface density is then $\rho_{s}=5.56\times 10^{18}$ molecules/m$^{2}$.\cite{ref41} The Raman cross section is $\sigma\approx 0.358 \times 10^{-30}$ cm$^{2}$/sr, which was obtained by applying the $\lambda_{S}^{-4}$ scaling to the original data of the 1,340 cm$^{-1}$ line. \cite{ref42} Single antenna extinction spectra $E(\lambda)$ (in dB) are calculated through $E(\lambda)=T_{ref}(\lambda)-T_{ant}(\lambda)$ in which $T_{ref}(\lambda)$ is the power transmission (in dB) through the reference waveguide and $T_{ant}(\lambda)$ the power transmission (in dB) of a waveguide functionalized with one antenna. Linear extinction spectra $e(\lambda)\stackrel{\Delta}{=}e_{\lambda}$ are then given by $e(\lambda)=10^{E(\lambda)/10}$. A field and index profile monitor are used to extract the local field $\left|\textbf{E}(\textbf{r},\lambda)\right|$ and index $n(\textbf{r})$ around the antenna. The single antenna conversion efficiency
\begin{equation}
P_{A}(\lambda_{P},\lambda_{S})=\frac{\rho_{s}\sigma}{2t_{NTP}}\frac{\int\int\int_{V_{m}} n_{g}(\lambda_{P})n_{g}(\lambda_{S})\lambda_{S}^{2}\left|\textbf{E}(\textbf{r},\lambda_{P})\right|^{2}\left|\textbf{E}(\textbf{r},\lambda_{S})\right|^{2}d\textbf{r}}{\left(\int\int n(\textbf{r})^{2}\left|\textbf{E}^{m}(\textbf{r},\lambda_{P})\right|^{2}d\textbf{r}\right)\left(\int\int n(\textbf{r})^{2}\left|\textbf{E}^{m}(\textbf{r},\lambda_{S})\right|^{2}d\textbf{r}\right)} \nonumber
\end{equation}
is calculated by integrating the local fields over the effective monolayer volume $V_{m}$ in which the index satisfies $n(\textbf{r})\left.\right|_{\textbf{r}\in V_{m}}=n_{NTP}$ (see Supplementary Information S3). The group index of the waveguide mode is $n_{g}(\lambda)$. The denominator is calculated using the modal fields $\textbf{E}^{m}(\textbf{r},\lambda)$ of a non-functionalized reference waveguide and the local field enhancement is given by the ratio of the local and modal electric fields: $\beta(\textbf{r},\lambda)=\frac{\left|\textbf{E}(\textbf{r},\lambda)\right|}{\left|\textbf{E}^{m}(\textbf{r},\lambda)\right|}$. At a certain position, the Raman enhancement factor $EF_{R}$ is calculated as $EF_{R}=\beta(\lambda_{P})^{2}\beta(\lambda_{S})^{2}$. Numerically calculated values are mentioned in the main text.

\section{Acknowledgement}
The authors acknowledge Josine Loo (imec) for performing the e-beam lithography, Liesbet Van Landschoot for making the SEM images, Ananth Subramanian for useful discussions and Stephane Clemmen for providing feedback on the manuscript. This research was funded by the ERC grant InSpectra. F.P acknowledges support from the Bijzonder Onderzoeksfonds (BOF) fellowship of Ghent University.

\section{Author Contributions}
F.P. fabricated the samples, performed the numerical simulations and experiments, analyzed the data, developed the analytical model and wrote the paper. A.D. built the experimental setup and developed the measurement procedure. P.V.D., N.L.T. and R.B. supervised the work. All authors provided feedback on the manuscript.

\section{Competing financial interests}
The authors declare no competing financial interests.

\newpage

\section{Supporting Information}

\subsection{S1: SERS measurement setup}
SERS spectra are measured with the setup depicted in Figure S1. A tunable Ti:saph laser is set to a pump wavelength of $\lambda_{P}=785$ nm (red) after which the polarized beam passes through a half-wave plate ($\lambda/2$) in order to rotate the polarization to a TE-polarized beam. A beamsplitter BS1 then splits the beam into two parts (solid and dashed red line). The solid path is used to generate the forward propagating Raman beam and passes through a laser line filter (LLF) at $\lambda_{P}$ for side-band suppression before it is coupled into the chip by an aspheric lens (ASPH). The output beam is then collected with an objective (OBJ) and passes through a polarizer P (set to TE) before it is filtered by a dichroic mirror which reflects the pump beam and transmits all Stokes wavelengths (green). The Stokes light is collected into a fiber using a parabolic mirror collimator (PMC) after which the fiber is split by a fiber splitter (FS) of which 1\% goes to a power meter (PM) and 99\% to a commercial spectrometer from ANDOR (Shamrock 303i spectrometer and iDus 416 cooled CCD camera). Mirror M2 blocks the second (dashed) path during the measurement but can be removed for alignment purposes. The camera (CAM) and the 1\% fiber tap are used during alignment and to measure the transmitted power $P_{T}$. In order to align the sample we initially set the wavelength of the Ti:saph to $\lambda_{T}=800$ nm (such that it can be transmitted through DM1 and collected in the power meter) and maximize the transmitted power $P_{T}$ in the forward path. After optimizing the transmission the laser is tuned back to $\lambda_{P}=785$ nm.
\begin{figure}[htbp]
  \centering
 	\includegraphics[trim= 4cm 6cm 0cm 4cm, clip=true,width=1\textwidth]{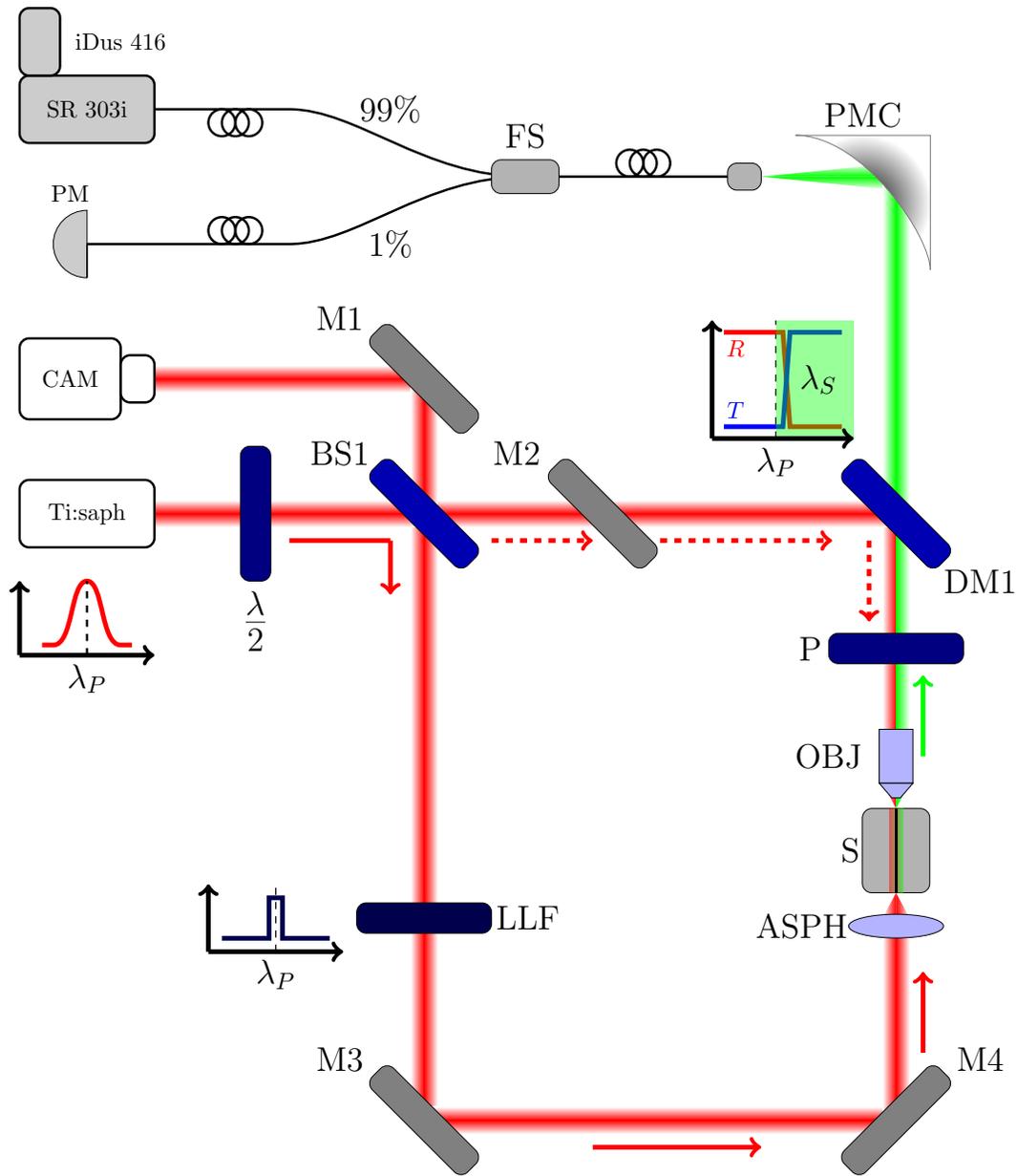}\label{figS1}
  \caption*{Supplementary Figure S1: \textbf{Measurement setup.} Ti:saph: tunable Ti:saphire laser emitting the pump beam at $\lambda_{P}=785$ nm, CAM: camera, PM: power meter, SR 303i and iDus 416: spectrometer and cooled CCD detector from ANDOR, BS1: beamsplitter, $\lambda/2$: half-wave plate, LLF: laser line filter for 785 nm, P: polarizer, M1/M3/M4: fixed mirrors, M2: removable mirror/beam block, OBJ: objective (50X, NA=0.9), ASPH: aspheric lens (NA=0.5), S: sample stage, DM1: dichroic mirror (reflection $R$ and transmission $T$ shown), PMC: parabolic mirror collimator (EFL=15 mm, NA=0.2), FS: fiber splitter.}
\end{figure}

\newpage

\subsection{S2: Extinction measurement}
Single antenna extinction spectra, resulting from the plasmon resonance, are measured with the setup depicted in Figure S2. Light from an NKT EXR-4 supercontinuum source (SC) is filtered through a near-IR filter (NIRF) to filter out the relevant wavelength region. Subsequently it is coupled in a fiber with a fiber coupling unit (FC). This fiber is plugged into a fiberbench consisting of 3 parts (fixed to the same bench): an achromatic fiber collimator (C) which converts the fiberized light to a free-space beam, a free-space broadband polarizer (P) which polarizes the unpolarized light into a TE-beam and an aspheric lens (ASPH) used to focus the free-space beam on the input facet of the chip. This fiberbench (marked by the light-blue area) is mounted on a piezo-controlled stage (XYZ) in order to precisely couple the supercontinuum light into the chip. At the output facet a lensed fiber (LF) is used to capture the transmitted light. This lensed fiber is also connected to a piezo-controller for accurate positioning. Finally the light is coupled to an optical spectrum analyzer (OSA) and the spectra are read out by Python controlled software (PC).
\begin{center}
\begin{figure}[htbp]
  \centering
 	\includegraphics[trim= 5cm 21cm 9cm 4cm, clip=true,width=1\textwidth]{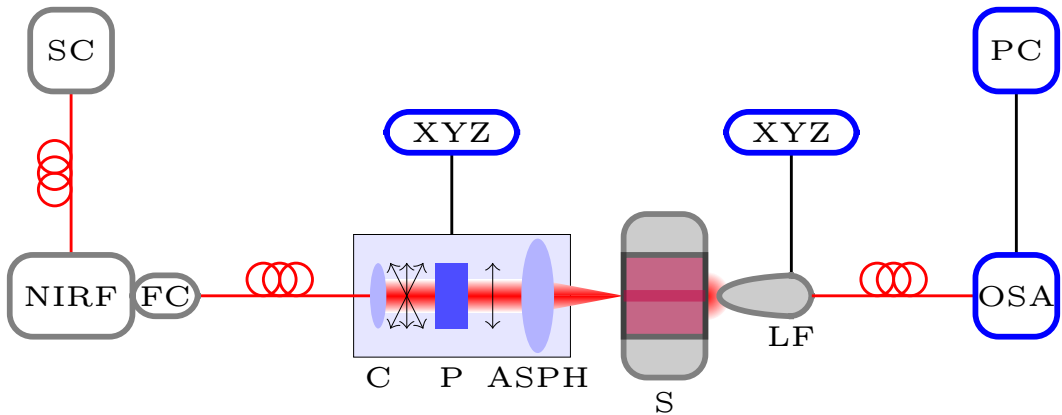}
  \caption*{Supplementary Figure S2: \textbf{Resonance measurement.} SC: supercontinuum source, NIRF: near-IR filter, FC: fiber coupling unit, C: achromatic fiber collimator, P: free-space broadband polarizer, ASPH: aspheric lens (NA=0.68), LF: lensed fiber, XYZ: piezo controller stage, S: sample stage, OSA: Optical Spectrum Analyzer, PC: OSA control using Python based measurement framework.}
\end{figure}
\end{center}
The single antenna extinction curves $E(\lambda)$ (in dB) can then be calculated through $E(\lambda)=(T_{ref}-T_{Nant})/N$ in which $T_{ref}$ is the power transmission (in dB) through the reference waveguide and $T_{Nant}$ the power transmission (in dB) of a waveguide functionalized with $N$ antennas. \cite{ref25} In Figure S3 the single antenna extinction curve of a waveguide functionalized with $N=40$ bowtie antennas ($\alpha\approx 60^{\circ}$, $H\approx 100$ nm and $\Delta\approx 40$ nm) is shown. The extinction exhibits periodic fringes on the broad envelope which are attributed to interference effects of the plasmonic array. On one hand, the array forms a multiple Fabry-Perot resonator of which the expected free spectral range FSR of 19.9 nm around 1,250 cm$^{-1}$ (using the fabricated array period of $\Lambda=10\mu$m) matches well with the experimentally obtained value ($\approx 20.8$ nm). On the other hand, the waveguide mode interferes with the radiative decay of the plasmon mode. This will affect the specific lineshape and strength of the fringes. The spectral positions at which we see sudden changes in the Raman background (see main text) coincide with the fringes observed on the extinction curves. Hence these features are not attributed to specific Raman lines.
\begin{center}
\begin{figure}[htbp]
  \centering
 	\includegraphics[trim= 3cm 10.5cm 3cm 10.5cm, clip=true,width=1\textwidth]{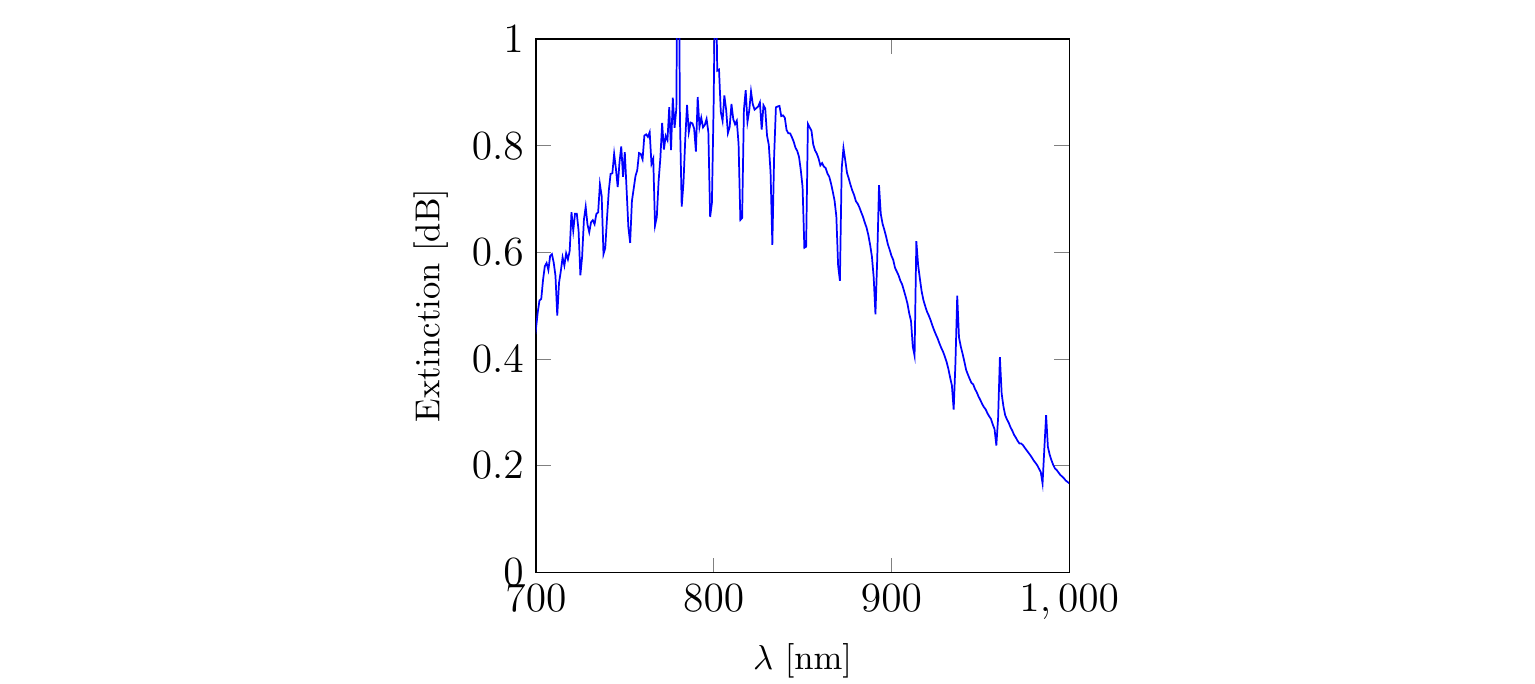}
  \caption*{Supplementary Figure S3: \textbf{Single antenna extinction spectrum.} Single antenna extinction spectrum of the waveguide functionalized with $N=40$ bowtie antennas ($\alpha\approx 60^{\circ}$, $H\approx 100$ nm and $\Delta\approx 40$ nm).}
\end{figure}
\end{center}

\newpage

\subsection{S3: Derivation of the analytical on-chip SERS model}
The power $P_{wg}(\textbf{r}_{0},\lambda)$ coupled into the forward propagating waveguide mode $\textbf{E}^{m}(\textbf{r},\lambda)$ as a result of a radiating dipole at position $\textbf{r}_{0}$ is given by
\begin{equation}
	\frac{P_{wg}(\textbf{r}_{0},\lambda)}{P_{0}}=\frac{3}{8\pi}\frac{n_{g}(\lambda)}{n_{m}}\left(\frac{\lambda}{n}\right)^{2}\frac{\epsilon_{o}\epsilon(\textbf{r}_{0})\left|\textbf{e}_{d}\cdot \textbf{E}^{m}(\textbf{r}_{0},\lambda)\right|^{2}}{\int\int_{A_{wg}} \epsilon_{o}\epsilon(\textbf{r})\left|\textbf{E}^{m}(\textbf{r},\lambda)\right|^{2}d\textbf{r}\left.\right|_{z=z_{0}}} \label{eq1}
\end{equation}
where $P_{0}=\frac{\omega^4\left|\textbf{d}_{0}\right|^{2}}{12\pi\epsilon_{0}c^{3}}$ is the power radiated in free space at a wavelength $\lambda=\frac{2\pi c}{\omega}$, $\epsilon(\textbf{r})$ the relative permittivity at position $\textbf{r}$, $n_{g}(\lambda)$ the group index of the waveguide mode, $n_{m}$ the refractive index of the medium in which the dipole is placed and $\textbf{e}_{d}$ the unit vector along the dipole moment vector $\textbf{d}_{0}$. \cite{ref39} The integral in the denominator is calculated over a waveguide cross-section $A_{wg}$ in the $xy$-plane and evaluated at the dipole position $z=z_{0}$ (coordinate axes are defined in Figure S4(a) and S4(b)).
\par
Functionalizing the waveguide with an antenna however introduces a perturbation to the modal field $\textbf{E}^{m}(\textbf{r},\lambda)$ of the waveguide. In order to investigate whether formula (1) is still valid when the waveguide is functionalized with a metallic antenna, two sets of simulations were performed . In the first set (Figure S4(a)) the coupling of dipole radiation into the fundamental TE-mode is investigated. This allows an explicit evaluation of the left-hand side of equation (1). Dipole sources with fixed dipole moment vector $\textbf{d}_{0}$ were placed at different positions $\textbf{r}_{0}$ around the antenna (shown as black dots in Figure S4(a)). In the second set (Figure S4(b)) the fundamental TE-mode is launched into the waveguide functionalized with the same antenna in order to calculate the fields at positions $\textbf{r}_{0}$ where the dipole sources in the first simulation set were located. In this way it is possible to calculate the right-hand side of equation (1). The integral in the denominator is approximated in all calculations by evaluating it on a reference waveguide without antenna, so using the true modal fields and not the perturbed fields. This approximation allows a sufficient accurate evaluation of the right-hand side of equation (1) as is confirmed in Figure S4(c) where the simulation results are depicted. The red curve represents an incoherent superposition of the power coupled into the TE-mode as a result of the complete set of dipole emitters. The dashed blue curve is the predicted power that couples into the TE-mode, and is calculated as an incoherent superposition of the predicted coupled powers for each dipole. It is clear that the predicted power coupled into the TE-mode (simulation set 2) matches very well with the explicit calculation using dipole emitters (simulation set 1).
\begin{figure}[htbp]
  \centering
 	\includegraphics[trim= 5.3cm 16cm 2.3cm 4.5cm, clip=true,width=1\textwidth]{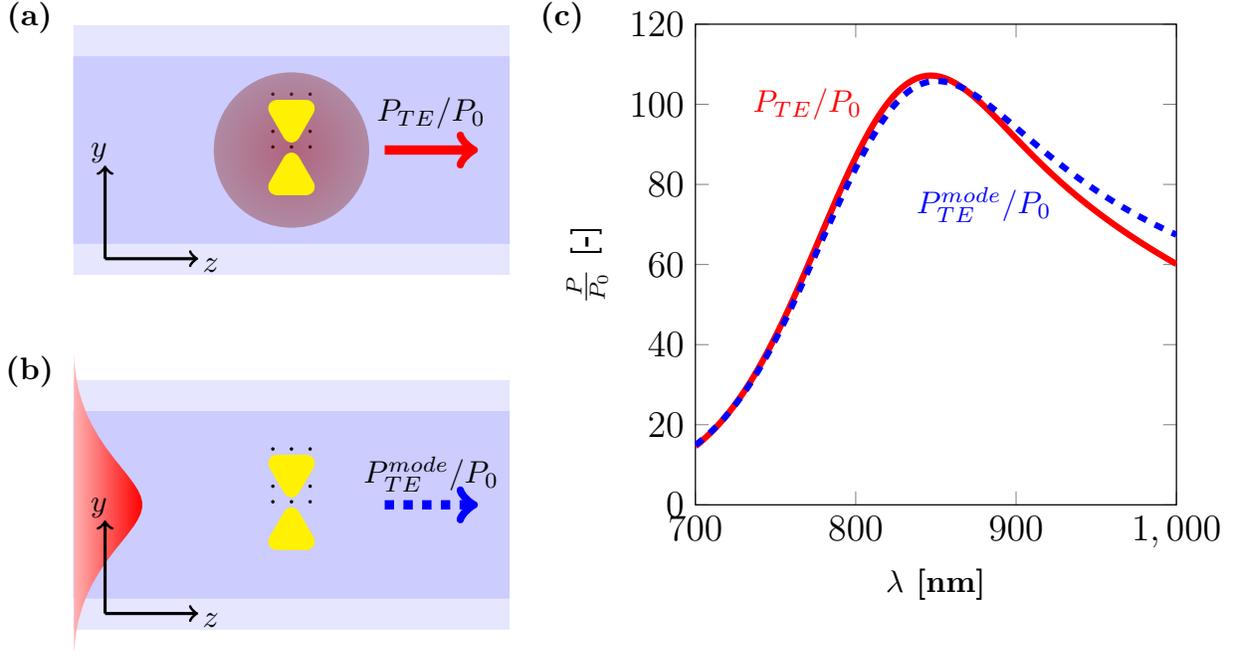}
  \caption*{Supplementary Figure S4: \textbf{Dipole radiation.} \textbf{(a)} Simulation set 1: collection of incoherent dipoles emitting into the fundamental TE-mode (red). \textbf{(b)} Simulation set 2: Calculation of the electric fields at the dipole positions of simulation set 1 using a fundamental TE-mode excitation. Based on these fields one can extract the predicted dipole emission (dashed blue). \textbf{(c)} Simulation results: incoherent superposition of the TE-coupled power due to all dipoles (red) and predicted TE-coupled power using the electric field values (dashed blue).}
\end{figure}
\par
Since the right-hand side of equation (1) faithfully represents the power coupled into the fundamental TE-mode, it is used to derive an analytical model predicting the total Stokes power coupled into the waveguide as a result of an array of $N$ antennas on top of the waveguide (see Figure 4(a) in the main text). First of all we write the dipole strength $\left|\textbf{d}_{0}\right|^2$ as a function of the guided pump power $P_{pump}$ in the waveguide
\begin{equation}
\left|\textbf{d}_{0}\right|^2 = \alpha_{m}^{2}\left|\textbf{E}_{d}^{m}(\textbf{r}_{0},\lambda)\right|^{2}\frac{n_{g}(\lambda)P_{pump}}{\int\int c\epsilon_{o}\epsilon(\textbf{r})\left|\textbf{E}^{m}(\textbf{r},\lambda)\right|^{2}d\textbf{r}\left.\right|_{z=z_{input}}}	\nonumber
\end{equation}
in which $\alpha_{m}$ is the molecular polarizability and $\textbf{E}_{d}^{m}(\textbf{r}_{0},\lambda)=\textbf{e}_{d}\cdot \textbf{E}^{m}(\textbf{r}_{0},\lambda)$ the field strength at the dipole position. The integral is evaluated at the input $z_{input}=0$ of the waveguide. \cite{ref36} The quantity
\begin{equation}
\eta_(\textbf{r}_{0},\lambda_{P},\lambda_{S})=\frac{n_{g}(\lambda_{P})n_{g}(\lambda_{S})\lambda_{S}^{2}}{n_{m}}\frac{\left|\textbf{E}_{d}^{m}(\textbf{r}_{0},\lambda_{P})\right|^{2}\left|\textbf{E}_{d}^{m}(\textbf{r}_{0},\lambda_{S})\right|^{2}}{\left(\int\int\epsilon(\textbf{r})\left|\textbf{E}^{m}(\textbf{r},\lambda_{P})\right|^{2}d\textbf{r}\right)\left(\int\int\epsilon(\textbf{r})\left|\textbf{E}^{m}(\textbf{r},\lambda_{S})\right|^{2}d\textbf{r}\right)}.	\label{eqeta}
\end{equation}
solely depends on the modal fields and can be evaluated on a reference waveguide. Both integrals in the denominator are evaluated along the same cross-sectional area of the reference waveguide (in all simulations $n_{m}=1$). The power $P_{wg}^{n}$ coupled into the fundamental TE-mode due to a collection of incoherently radiating dipoles around antenna $n$ ($n=1\ldots N$) is then approximately given by
\begin{equation}	\frac{P_{wg}^{n}}{P_{pump}}\approx \frac{\rho\sigma(\lambda_{S})}{2}\text{e}^{-\alpha(\lambda_{P})(L_{1}+(n-1)\Lambda)}e(\lambda_{P})^{1-n}\int\int\int_{V_{m}}\eta_(\textbf{r}_{0},\lambda_{P},\lambda_{S})\beta(\textbf{r}_{0},\lambda_{P})^{2}\beta(\textbf{r}_{0},\lambda_{S})^{2}d\textbf{r}_{0} \label{eq2}
\end{equation}	
in which $\rho$ is the molecular density, $\sigma(\lambda_{S})=\pi^{2}\alpha_{m}^{2}/(\epsilon_{0}^{2}\lambda_{S}^{4})$ the Raman cross section\cite{ref40}, $\alpha(\lambda_{P})=\alpha_{P}$ the waveguide loss at the pump wavelength, $e(\lambda_{P})=e_{P}$ the linear antenna extinction at the pump wavelength, $V_{m}$ the volume in which the molecules are situated and $\beta(\textbf{r}_{0},\lambda_{P/S})=\beta_{P/S}(\textbf{r}_{0})$ the field enhancement factor, at the dipole position $\textbf{r}_{0}$, at the pump ($P$) and Stokes ($S$) wavelength respectively:
\begin{equation}
	\beta(\textbf{r}_{0},\lambda)=\frac{\left|\textbf{E}_{d}^{ant}(\textbf{r}_{0},\lambda)\right|}{\left|\textbf{E}_{d}^{m}(\textbf{r}_{0},\lambda)\right|} \nonumber
\end{equation}
whereby $\textbf{E}_{d}^{ant}(\textbf{r}_{0},\lambda)$ is the local field around the antenna surface. For notational simplicity we rewrite formula (3) as
\begin{equation}	
\frac{P_{wg}^{n}}{P_{pump}}\approx P_{A}(\lambda_{P},\lambda_{S})\text{e}^{-\alpha_{P}(L_{1}+(n-1)\Lambda)}e_{P}^{1-n} \nonumber
\end{equation}
in which
\begin{equation}
	P_{A}(\lambda_{P},\lambda_{S})=\frac{\rho\sigma(\lambda_{S})}{2}\int\int\int_{V_{m}}\eta_(\textbf{r}_{0},\lambda_{P},\lambda_{S})\beta(\textbf{r}_{0},\lambda_{P})^{2}\beta(\textbf{r}_{0},\lambda_{S})^{2}d\textbf{r}_{0} \nonumber
\end{equation}
is a dimensionless antenna dependent factor which incorporates the specific field enhancement profile near the antenna surface. Formula (3) takes into account the waveguide loss and the loss induced by the antennas in front of the $n^{\text{th}}$ antenna as a result of which the actual excitation power decays. The signal (at Stokes wavelength $\lambda_{S}$) still has to pass $N-n$ antennas and propagate along a distance $L_{1}+(N-n)\Lambda$ such that the power $P_{wg,out}^{n}$ reaching the output of the waveguide is given by
\begin{align}
\frac{P_{wg,out}^{n}}{P_{pump}}&\approx P_{A}(\lambda_{P},\lambda_{S})\left(\text{e}^{-\alpha_{P}(L_{1}+(n-1)\Lambda)}e_{P}^{1-n}\right)\left(\text{e}^{-\alpha_{S}(L_{1}+(N-n)\Lambda)}e_{S}^{n-N}\right). \nonumber \\
&=P_{A}(\lambda_{P},\lambda_{S})\left(\text{e}^{-\alpha_{P}(L_{1}-\Lambda)-\alpha_{S}(L_{1}+N\Lambda)}e_{P}e_{S}^{-N}\right)\left(\left(\frac{e_{S}}{e_{P}}\right)\text{e}^{(\alpha_{S}-\alpha_{P})\Lambda}\right)^{n}. \nonumber		
\end{align}
Since $N$ antennas contribute incoherently to the signal, the total amount of Stokes light at the output of the waveguide is given by
\begin{equation}
\frac{P_{wg,out}^{tot}}{P_{pump}}\approx P_{A}(\lambda_{P},\lambda_{S})\left(\text{e}^{-\alpha_{P}(L_{1}-\Lambda)-\alpha_{S}(L_{1}+N\Lambda)}e_{P}e_{S}^{-N}\right)\sum_{n=1}^{N}\left(\left(\frac{e_{S}}{e_{P}}\right)\text{e}^{(\alpha_{S}-\alpha_{P})\Lambda}\right)^{n}. \nonumber	
\end{equation}
For the considered waveguide platform it is reasonable to approximate $\alpha_{P}\approx\alpha_{S}=\alpha_{wg}$ such that the forward propagating Raman power is given by
\begin{align}
\frac{P_{wg,out}^{tot}}{P_{pump}}
&=\frac{P_{tot}}{P_{pump}} \nonumber \\
&\approx P_{A}(\lambda_{P},\lambda_{S})\left(\text{e}^{-\alpha_{wg}(2L_{1}+(N-1)\Lambda)}e_{P}e_{S}^{-N}\right)\sum_{n=1}^{N}\left(\frac{e_{S}}{e_{P}}\right)^{n} \nonumber	\\
&\approx P_{A}(\lambda_{P},\lambda_{S})\left(\text{e}^{-2\alpha_{wg}L_{1}}e_{S}^{1-N}\right)\left(\frac{1-\left(\frac{e_{S}}{e_{P}}\right)^{N}}{1-\left(\frac{e_{S}}{e_{P}}\right)}\right)= FOM(N,\lambda_{P},\lambda_{S})\text{e}^{-2\alpha_{wg}L_{1}}, \nonumber
\end{align}
where $L=2L_{1}+(N-1)\Lambda \approx 2L_{1}$ has been used. This result applies to Stokes light co-propagating with the pump beam. For a fixed antenna geometry and $\lambda_{P}$ and $\lambda_{S}$, the optimum number of antennas $N_{opt}$ that should be patterned on a waveguide to maximize the SERS signal is given by
\begin{equation}
	N_{opt}=\frac{\log\left(\frac{\log\left(e_{S}\right)}{\log\left(e_{P}\right)}\right)}{\log\left(\frac{e_{S}}{e_{P}}\right)}. \nonumber
\end{equation}
For the specific case of a monolayer, one has a molecular surface density $\rho_{s}$ (number of molecules per $m^{2}$) rather than a volume density $\rho$ (number of molecules per $m^{3}$). In this limiting case the single antenna conversion efficiency is theoretically defined by
\begin{equation}
	P_{A}(\lambda_{P},\lambda_{S})=\frac{\rho_{s}\sigma(\lambda_{S})}{2}\int\int_{A_{m}}\eta_(\textbf{r}_{0},\lambda_{P},\lambda_{S})\beta(\textbf{r}_{0},\lambda_{P})^{2}\beta(\textbf{r}_{0},\lambda_{S})^{2}d\textbf{r}_{0} \nonumber
\end{equation}
where the integration is now performed over the surface area $A_{m}$ covered by the monolayer. Numerically one however always needs to model the presence of such a monolayer by introducing a finite thickness $t_{m}$. As a result an effective monolayer volume $V_{m}\approx t_{m}\times A_{m}$ is defined. For any numerical evaluation the fields are hence integrated over this volume $V_{m}$ such that
\begin{equation}
	\int\int\int_{V_{m}}\eta_(\textbf{r}_{0},\lambda_{P},\lambda_{S})\beta(\textbf{r}_{0},\lambda_{P})^{2}\beta(\textbf{r}_{0},\lambda_{S})^{2}d\textbf{r}_{0}\approx t_{m}\times\int\int_{A_{m}}\eta_(\textbf{r}_{0},\lambda_{P},\lambda_{S})\beta(\textbf{r}_{0},\lambda_{P})^{2}\beta(\textbf{r}_{0},\lambda_{S})^{2}d\textbf{r}_{0}\nonumber
\end{equation}
The single antenna conversion efficiency is then given by
\begin{equation}
	P_{A}(\lambda_{P},\lambda_{S})\approx\frac{\rho_{s}\sigma(\lambda_{S})}{2t_{m}}\int\int\int_{V_{m}}\eta_(\textbf{r}_{0},\lambda_{P},\lambda_{S})\beta(\textbf{r}_{0},\lambda_{P})^{2}\beta(\textbf{r}_{0},\lambda_{S})^{2}d\textbf{r}_{0} \nonumber
\end{equation}
From a numerical point of view, an effective volume density $\rho_{eff}=\rho_{s}/t_{m}$ is introduced which is then multiplied with the integral over the effective monolayer volume $V_{m}$.
\newpage
In order to correlate the Stokes power at the output facet of the chip and the actual detected signal counts $C_{S}$, normalized with the integration time $T$, we need to take into account the characteristics of all optics in our setup. The Stokes power reaching the detector surface is given by
\begin{align}
	P_{S} &= \gamma^{out}\times T_{out} \times P_{tot} = \gamma^{out}\times T_{out} \times \left(FOM\text{e}^{-\alpha_{wg}L}P_{pump}\right) \nonumber \\
	&= \gamma^{out}\times T_{out} \times \left(FOM\text{e}^{-\alpha_{wg}L}\left(\gamma^{in}P_{in}\right)\right) \nonumber \\
	&= FOM  \left(\gamma^{in}\gamma^{out}P_{in}\text{e}^{-\alpha_{wg}L}\right) \times T_{out} \nonumber
\end{align}
in which $\gamma^{in}$ and $\gamma^{out}$ are the coupling efficiencies in and out of the chip respectively, $P_{in}$ is the power just before the input facet of the chip and $T_{out}$ is the transmission through all optics between the output of the chip and the detector surface. The quantity $\gamma^{in}\gamma^{out}P_{in}\text{e}^{-\alpha_{wg}L}$ can be written as
\begin{equation}
	\gamma^{in}\gamma^{out}P_{in}\text{e}^{-\alpha_{wg}L}=P_{T}/T_{PM} \nonumber
\end{equation}
in which $P_{T}$ is the transmitted power as measured by the power meter (see Supplementary Information S1) and $T_{PM}$ is the optical transmission between the output of the chip and the input of the powermeter. This transmission is related to $T_{out}$ by $T_{out}\approx 100 \times T_{PM} \times T_{spec}$ in which $T_{spec}$ is the transmission between the input slit of the spectrometer and the detector surface. The factor 100 stems from the fact that the fiber splitter only transmits 1\% of the power to the power meter. So eventually $P_{S}$ can be written as $P_{S}=FOM \times P_{T} \times 100 T_{spec}$.
By using the spectrometer sensitivity $\chi$ (defined as the number of electrons per count) and quantum efficiency $QE(\lambda_{S})$ (number of electrons per number of incident photons) $P_{S}$ can be related to the measured signal counts $C_{S}$ per unit integration time $T$ by
\begin{equation}
	P_{S}=\left(\frac{C_{S}}{T}\right)\left(\frac{hc}{\lambda_{S}}\right)\left(\frac{\chi}{QE(\lambda_{S})}\right). \nonumber
\end{equation}
Finally one gets
\begin{equation}
	\frac{C_{S}}{T}=FOM \times 100T_{spec}\left(\frac{\lambda_{S}}{hc}\right)\left(\frac{QE(\lambda_{S})}{\chi}\right)P_{T}= FOM \times C^{*}(\lambda_{S}). \nonumber
\end{equation}
The transmitted power $100P_{T}$ is about 1mW for the given measurement setup. If we consider the 1340 cm$^{-1}$ line then $\frac{QE(\lambda_{S})}{\chi}\approx 1$ counts/photon and $T_{spec}\approx0.44$ (using data supplied by the manufacturer) such that $C^{*}\approx 1.94\times 10^{15}$ counts/sec. $C^{*}(\lambda_{S})$ can now be used as a conversion factor when the experimental signal counts are fitted to the analytical model (and hence the figure of merit). In this way one can rigorously quantify the parameters $e_{P}$, $e_{S}$ and $P_{A}$ and compare them with the theoretical predictions.
\newpage
	
\subsection{S4: Derivation of the background signal}
Similar reasonings as in the Supplementary Information S3 can be applied to derive the total background signal generated along the waveguide. This background mostly stems from the SiN core and is hence generated when the pump light propagates along the total length $L$ of the waveguide. For a functionalized waveguide the attenuation of the pump and Stokes light due to the antenna array also needs to be incorporated. The dipoles which give rise to the background are mainly situated in the core, so for a given core cross-sectional area $A_{core}$ equation (2) has to be integrated over the complete waveguide core. This quantity is defined as
\begin{equation}
\eta_{core}(\lambda_{P},\lambda_{S})=\eta_{c}=\int\int_{A_{core}}\eta_(\textbf{r}_{0},\lambda_{P},\lambda_{S})d\textbf{r}_{0}. \nonumber
\end{equation}
For a given molecular core density $\rho_{c}$ and core scattering cross section $\sigma_{c}$, the total background $P_{bg}$ propagating in the forward direction is then calculated through
\begin{align}
	\frac{P_{bg}}{P_{pump}}=&\frac{\rho_{c}\sigma_{c}\eta_{c}}{2}\left\{\left(\int_{0}^{L_{1}}\text{e}^{-\alpha_{P}z}\text{e}^{-\alpha_{S}(L-z)}dz\right)e_{S}^{-N}+\left(\int_{L_{1}+(N-1)\Lambda}^{2L_{1}+(N-1)\Lambda}\text{e}^{-\alpha_{P}z}\text{e}^{-\alpha_{S}(L-z)}dz\right)e_{P}^{-N}\right.+ \nonumber \\ 
	&\left.\sum_{n=0}^{N-2}\left(\int_{L_{1}+n\Lambda}^{L_{1}+(n+1)\Lambda}\text{e}^{-\alpha_{P}z}\text{e}^{-\alpha_{S}(L-z)}dz\right)e_{P}^{-(n+1)}e_{S}^{-(N-n-1)}\right\}. \nonumber
\end{align}
After some calculation and using $\alpha_{P}\approx\alpha_{S}=\alpha_{wg}$ and $L\approx L_{1}/2$ ($2L_{1}>>(N-1)\Lambda$), the total forward propagating background is given by
\begin{align}
\frac{P_{bg}}{P_{pump}}&=\frac{\rho_{c}\sigma_{c}\eta_{c}}{2}\text{e}^{-\alpha_{wg}L}\left\{(e_{P}^{-N}+e_{S}^{-N})\frac{L}{2}+\Lambda\frac{e_{S}^{1-N}}{e_{P}}\left(\frac{1-\left(\frac{e_{S}}{e_{P}}\right)^{N-1}}{1-\left(\frac{e_{S}}{e_{P}}\right)}\right)\right\} \nonumber \\
&\approx\frac{\rho_{c}\sigma_{c}\eta_{c}}{2}\text{e}^{-\alpha_{wg}L}\left\{(e_{P}^{-N}+e_{S}^{-N})\frac{L}{2}\right\} \nonumber \\
&=P_{B}(\lambda_{P},\lambda_{S})\text{e}^{-\alpha_{wg}L}\left(e_{P}^{-N}+e_{S}^{-N}\right)\approx P_{B}(\lambda_{P},\lambda_{S})\text{e}^{-2\alpha_{wg}L_{1}}\left(e_{P}^{-N}+e_{S}^{-N}\right).\nonumber
\end{align}

\newpage

\subsection{S5: Randomized fit to a generalized model}
The model derived in the Supplementary Information S3 and S4 assumes identical antennas. Due to fabrication errors there will always be differences among each of the antennas in the array. These differences have an impact on both the extinction $e_{P}$ and $e_{S}$ as well as on the single antenna conversion efficiency $P_{A}(\lambda_{P},\lambda_{S})$. While the experimental data described in this Letter can be fitted well to the ideal model, the fit is not perfect. In this section a randomized fit model that takes into account the potential deviations among different antennas is outlined. The power generated by antenna $n$ is given by:
\begin{equation}	
\frac{P_{wg}^{n}}{P_{pump}}\approx P_{A}(\lambda_{P},\lambda_{S})\text{e}^{-\alpha_{P}(L_{1}+(n-1)\Lambda)}e_{P}^{1-n}. \nonumber
\end{equation}
Since $L_{1}>>N\Lambda$ this can be simplified to
\begin{equation}	
\frac{P_{wg}^{n}}{P_{pump}}\approx P_{A}(\lambda_{P},\lambda_{S})\text{e}^{-\alpha_{P}L_{1}}e_{P}^{1-n}. \nonumber
\end{equation}
The formula however assumes a constant pump extinction from the previous $n-1$ antennas. Furthermore the factor $P_{A}(\lambda_{P},\lambda_{S})$ is also assumed constant for each antenna. If we allow that each of the antennas has a different extinction $e_{P}^{m}$ and antenna dependent factor $P_{A}^{m}(\lambda_{P},\lambda_{S})$ ($m=1\ldots N$), then the power generated by antenna $n$ is given by
\begin{equation}	
\frac{P_{wg}^{n}}{P_{pump}\text{e}^{-\alpha_{P}L_{1}}}\approx P_{A}^{n}(\lambda_{P},\lambda_{S})\prod_{m=1}^{n-1}(e_{P}^{m})^{-1}. \nonumber
\end{equation}
Applying a similar reasoning to the Stokes light that has to propagate along $N-n$ other antennas (and a length $L_{1}$), the Stokes power (generated by antenna $n$) reaching the output is given by
\begin{equation}	
\frac{P_{wg,out}^{n}}{P_{pump}\text{e}^{-\alpha_{P}2L_{1}}}\approx P_{A}^{n}(\lambda_{P},\lambda_{S})\left(\prod_{m=1}^{n-1}(e_{P}^{m})^{-1}\right)\left(\prod_{m=n+1}^{N}(e_{S}^{m})^{-1}\right). \nonumber
\end{equation}
The total generated Stokes power from all $N$ (potentially different) antennas is then finally
\begin{equation}
\frac{P_{tot}}{P_{pump}\text{e}^{-2\alpha_{wg}L_{1}}}\approx \sum_{n=1}^{N} P_{A}^{n}(\lambda_{P},\lambda_{S})\left(\prod_{m=1}^{n-1}(e_{P}^{m})^{-1}\right)\left(\prod_{m=n+1}^{N}(e_{S}^{m})^{-1}\right). \label{RandTotSignal}
\end{equation}
For the background one can analogously write
\begin{equation}
\frac{P_{bg}}{P_{pump}\text{e}^{-2\alpha_{wg}L_{1}}}\approx P_{B}\left(\prod_{m=1}^{N}(e_{P}^{m})^{-1}+\prod_{m=1}^{N}(e_{S}^{m})^{-1}\right) \label{RandTotNoise}
\end{equation}
In this case $P_{B}$ is a constant since it only depends on the waveguide parameters. The shot noise associated to the background is then simply proportional to $\sqrt{P_{bg}}$. The number of counts/sec can be obtained by multiplying equations (4) and (5) with $C^{*}(\lambda_{S})$ (see Supplementary Information S3). Based on the conversion factor $C^{*}(\lambda_{S})$ and a fit to the experimental data, one can quantitatively determine a value for the parameters in the model. For a fixed $\lambda_{P}$ and $\lambda_{S}$, $P_{A}(\lambda_{P},\lambda_{S})C^{*}(\lambda_{S})$ will be denoted by $P_{A}^{*}$.
\par
In order to make a randomized fit of the signal and shot noise at the same time, $N$ random numbers (based on a normal distribution) are independently generated for the pump and Stokes extinction and for the single antenna conversion efficiency (so three random numbers are generated for each antenna in the array). Using these random numbers, formulas (4) and (5) are evaluated. This process is repeated 1000 times in order to generate a statistically relevant distribution of the possible signal and shot noise counts. Ultimately the mean value $\mu$ and standard deviation $\sigma$ from the obtained signal and shot noise distributions are extracted for each number of antennas $N$. The $3\sigma$-intervals for both distributions ($[\mu-3\sigma,\mu+3\sigma]$) define an area that marks the possible signal and noise counts for a given uncertainty on the antenna parameters (according to formulas (4) and (5)). These areas are plotted in Figure 4(b) of the main text.
\par
The mean value of each of the three normal distributions (for $e_{P}$, $e_{S}$ and $P_{A}$) is determined by a fit of the experimental data to the ideal model. One could obtain a perfect fit to either the signal or shot noise data by fitting only one of the two equations ((4) or (5)). The fitting parameters would however not generate a perfect fit to the other non-fitted equation since both equations depend on $e_{P}$ and $e_{S}$. Therefore a constrained fit, in which one minimizes the fitting error to both equations simultaneously, is performed such that both the signal and shot noise data are represented well using the fitted values for $e_{P}$ and $e_{S}$. The values $e_{P}$, $e_{S}$ and $P_{A}^{*}$ (mind that $P_{A}^{*}$ and not $P_{A}$ is used because we fit the counts/sec) obtained from the constrained fit are then chosen to be the mean values ($\mu_{e_{P}}$, $\mu_{e_{S}}$ and $\mu_{P_{A}^{*}}$) of their respective normal distributions. The standard deviation $\sigma_{x}$ is chosen such that the $3\sigma_{x}$ intervals represent realistic deviations from the mean value ($x$ denotes one the three parameters). For the extinction this means e.g. that $\mu_{e_{P/S}}-3\sigma_{e_{P/S}}\geq 1$ (since 1 is the lower boundary for the linear extinction). Since $P_{A}^{*}$ is always positive, $\mu_{P_{A}^{*}}-3\sigma_{P_{A}^{*}}>0$ should also be satisfied. For the experimental data shown in Figure 3 of the main text, $\mu_{e_{P}}=1.1182$, $\mu_{e_{S}}=1.0845$ and $\mu_{P_{A}^{*}}=5.0394$ has been derived from the constrained fit. The sigma value for the extinction is chosen to be $\sigma=0.028$ such that $E_{P}\approx 0.49\pm 0.11$ dB and $E_{S}\approx 0.35\pm 0.11$ dB. For $P_{A}^{*}$ we similarly get $P_{A}^{*}\approx 5.04 \pm 1.5$ such that the single antenna conversion efficiency for the 1340 cm$^{-1}$ line is $P_{A}=(2.60 \pm 0.77) \times 10^{-15}$. All of these values satisfy the constraints mentioned above and represent realistic deviations of the three fitting parameters.


\newpage

\begin{thebibliography}{50}

\bibitem{ref1} Anker, J. N., et al. Biosensing with plasmonic nanosensors. \textit{Nat. Mater.} \textbf{7,} 442--453 (2008).
\bibitem{ref2} Willets, K. A. \& Van Duyne, R.P. Localized surface plasmon resonance spectroscopy and sensing. \textit{Annu. Rev. Phys. Chem.} \textbf{58,} 267--297 (2007).
\bibitem{ref3} Halas, N. J., Lal, S., Chang, W-S., Link, S. \& Nordlander, P. Plasmons in strongly coupled metallic nanostructures. \textit{Chem. Rev.} \textbf{111,} 3913--3961 (2011).


\bibitem{ref8} Giannini, V., Fern\'andez-Dom\'inguez, A. I., Heck, S. C. \& Maier, S. A. Plasmonic nanoantennas: fundamentals and their use in controlling the radiative properties of nanoemitters. \textit{Chem. Rev.} \textbf{111,} 3888--3912 (2011).



\bibitem{ref17} Chu, Y., Banaee, M. G. \& Crozier, K. B. Double-resonance plasmon substrates for surface-enhanced Raman scattering with enhancement at excitation and Stokes frequencies. \textit{ACS Nano} \textbf{4 (5),} 2804--2810 (2010).
\bibitem{ref18} McFarland, A. D., Young, M. A., Dieringer, J. A. \& Van Duyne, R. P. Wavelength-scanned surface-enhanced Raman excitation spectroscopy. \textit{J. Phys. Chem. B} \textbf{109,} 11279--11285 (2005).
\bibitem{ref19} Ye, J., et al. Plasmonic Nanoclusters: Near Field Properties of the Fano Resonance Interrogated with SERS. \textit{Nano Lett.} \textbf{12 (3),} 1660--1667 (2012).
\bibitem{ref20} Kasera, S., Biedermann, F., Baumberg, J.J., Scherman, O.A. \& Mahajan, S. Quantitative SERS Using the Sequestration of Small Molecules Inside Precise Plasmonic Nanoconstructs. \textit{Nano Lett.} \textbf{12 (11),} 5924--5928 (2012).
\bibitem{ref21} Gallinet, B., Siegfried, T., Sigg, H., Nordlander, P. \& Martin, O.J.F. Plasmonic Radiance: Probing Structure at the Angstr\"om Scale with Visible Light. \textit{Nano Lett.} \textbf{13 (2),} 497--503 (2013).
\bibitem{ref22} Siegfried, T., Ekinci, Y., Martin, O.J.F. \& Sigg, H. Gap Plasmons and Near-Field Enhancement in Closely Packed Sub-
10 nm Gap Resonators. \textit{Nano Lett.} \textbf{13 (11),} 5449--5453 (2013).
\bibitem{ref23} Li, J., et al. 300 mm Wafer-level, ultra-dense arrays of Au-capped nanopillars with sub-10 nm gaps as reliable SERS substrates. \textit{Nanoscale} \textbf{6,} 12391--12396 (2014).
\bibitem{ref24} Seok, T. J., Jamshidi, A., Eggleston, M. \& Wu, M.C. Mass-producible and efficient optical antennas with CMOS-fabricated nanometer-scale gap. \textit{Opt. Express} \textbf{21 (14),} 16561--16569 (2013).

\bibitem{ref32} Lin, S., Zhu, W., Jin, Y. \& Crozier, K.B. Surface-Enhanced Raman Scattering with Ag Nanoparticles Optically Trapped by a Photonic Crystal Cavity. \textit{Nano Lett.} \textbf{13 (2),} 559–-563 (2013).
\bibitem{ref33} Kong, L., Lee, C., Earhart, C.M., Cordovez, B. \& Chan, J.W. A nanotweezer system for evanescent wave excited surface enhanced Raman spectroscopy (SERS) of single nanoparticles. \textit{Opt. Express} \textbf{23 (5),} 6793–-6802 (2015).
\bibitem{ref44} Measor, P., et al. On-chip surface-enhanced Raman scattering detection using integrated liquid-core waveguides. \textit{Appl. Phys. Lett.} \textbf{90,} 211107-1–-211107-3 (2007).


\bibitem{ref25} Peyskens, F., et al. Bright and dark plasmon resonances of nanoplasmonic antennas evanescently coupled with a silicon nitride waveguide. \textit{Opt. Express} \textbf{23 (3),} 3088--3101 (2015).
\bibitem{ref26} Arnaud, L., et al. Waveguide-coupled nanowire as an optical antenna. \textit{J. Opt. Soc. Am. A} \textbf{30 (11),} 2347--2355 (2013).
\bibitem{ref27} Arango, F. B., Kwadrin, A. \& Koenderink, A. F. Plasmonic Antennas Hybridized with Dielectric Waveguides. \textit{ACS Nano} \textbf{6 (11),} 10156--10167 (2012).
\bibitem{ref28} F\'evrier, M., et al. Giant Coupling Effect between Metal Nanoparticle Chain and Optical Waveguide. \textit{Nano Lett.} \textbf{12,} 1032--1037 (2012).
\bibitem{ref29} F\'evrier, M., Gogol, P.,  Lourtioz, J-M. \& Dagens, B. Metallic nanoparticle chains on dielectric waveguides: coupled and uncoupled situations compared. \textit{Opt. Express} \textbf{21 (21),} 24504--24513 (2013).
\bibitem{ref30} F\'evrier, M., et al. Integration of short gold nanoparticles chain on SOI waveguide toward compact integrated bio-sensors. \textit{Opt. Express} \textbf{20 (16),} 17402--17410 (2012).
\bibitem{ref31} Chamanzar, M., Xia, Z., Yegnanarayanan, S. \& Adibi, A. Hybrid integrated plasmonic-photonic waveguides for on-chip localized surface
plasmon resonance (LSPR) sensing and spectroscopy. \textit{Opt. Express} \textbf{21 (26),} 32086--32098 (2013).

\bibitem{ref41} Mahmoud, M.A. Surface-Enhanced Raman Spectroscopy of Double-Shell Hollow Nanoparticles: Electromagnetic and Chemical Enhancements. \textit{Langmuir} \textbf{29,} 6253--6261 (2013).


\bibitem{ref36} Dhakal, A., et al. Evanescent excitation and collection of spontaneous Raman spectra using silicon nitride nanophotonic waveguides. \textit{Opt. Lett.} \textbf{39 (13),} 4025--4028 (2014).


\bibitem{ref37} Johnson, P.B., Christy, R.W. Optical Constants of the Noble Metals. \textit{Phys. Rev. B} \textbf{6 (12),} 4370--4379 (1972).
\bibitem{ref38} Haynes, W. M., Eds. \textit{CRC Handbook Of Chemistry And Physics} 95th Edition (Internet Version 2015) (CRC/Taylor and Francis, 2015).

\bibitem{ref42} Thomas, M., et al. Distinguishing chemical and electromagnetic enhancement in surface-enhanced Raman spectra: The case of para-nitrothiophenol. \textit{J. Raman Spectrosc.} \textbf{44,} 1497–-1505 (2013).

\bibitem{ref46} Martens, D., et al. Compact silicon nitride arrayed waveguide gratings for very near-infrared wavelengths. \textit{IEEE Photon. Technol. Lett.} \textbf{27(2),} 137–-140 (2015).

\bibitem{ref43} Pelton, M. Modified spontaneous emission in nanophotonic structures. \textit{Nature Photon.} \textbf{9,} 427–-435 (2015).
\bibitem{ref45} Tame, M.S., et al. Quantum plasmonics. \textit{Nature Phys.} \textbf{9,} 329-340 (2013).


\bibitem{ref39} Chul Jun, Y., Briggs, R.M., Atwater, H.A. \& Brongersma, M.L. Broadband enhancement of light emission in silicon slot waveguides. \textit{Opt. Express} \textbf{17 (9),} 7479--7490 (2009).

\bibitem{ref40} Long, D.A. \textit{The Raman Effect: A Unified Treatment Of The Theory Of Raman Scattering By Molecules} pp 96-97 (John Wiley \& Sons Ltd, Chichester, England, 2002).

\end{thebibliography}
\end{document}